\documentclass[%
 aip,apl
 amsmath,amssymb,
 reprint,%
]{revtex4-1}

\usepackage{graphicx}
\usepackage{epstopdf}

\usepackage{dcolumn}
\usepackage{bm}
\usepackage{amsmath}
\usepackage{color} 
\usepackage{siunitx}
\usepackage[mathlines]{lineno}
\usepackage{hyperref}

\begin{document}

\preprint{AIP/123-QED}

\title{Design and low-temperature characterization of a tunable microcavity for diamond-based quantum networks}

\author{Stefan Bogdanovi\'{c}}
\affiliation{QuTech, Delft University of Technology, PO Box 5046, 2600 GA Delft, The Netherlands}
\author{Suzanne B. van Dam}%
\affiliation{QuTech, Delft University of Technology, PO Box 5046, 2600 GA Delft, The Netherlands}
\author{Cristian Bonato}
\affiliation{QuTech, Delft University of Technology, PO Box 5046, 2600 GA Delft, The Netherlands}
\author{Lisanne C. Coenen}
\affiliation{QuTech, Delft University of Technology, PO Box 5046, 2600 GA Delft, The Netherlands}
\author{Anne-Marije J. Zwerver}
\affiliation{QuTech, Delft University of Technology, PO Box 5046, 2600 GA Delft, The Netherlands}
\author{Bas Hensen}
\affiliation{QuTech, Delft University of Technology, PO Box 5046, 2600 GA Delft, The Netherlands}
\author{Madelaine S.Z. Liddy}
\affiliation{Institute for Quantum Computing, University of Waterloo, Waterloo, ON, N2L3G1, Canada}
\author{Thomas Fink}
\affiliation{Institute of Quantum Electronics, ETH, CH-8093 Z\"urich, Switzerland}
\author{Andreas Reiserer}
\affiliation{QuTech, Delft University of Technology, PO Box 5046, 2600 GA Delft, The Netherlands}
\author{Marko Lon\v{c}ar}
\affiliation{John A. Paulson School of Engineering and Applied Sciences, Harvard University, Cambridge, Massachusetts 02138, USA}
\author{Ronald Hanson}
\affiliation{QuTech, Delft University of Technology, PO Box 5046, 2600 GA Delft, The Netherlands}
\email{r.hanson@tudelft.nl}

\date{\today}

\begin{abstract}
We report on the fabrication and characterization of a Fabry-Perot microcavity enclosing a thin diamond membrane at cryogenic temperatures. The cavity is designed to enhance resonant emission of single nitrogen-vacancy centers by allowing spectral and spatial tuning while preserving the optical properties observed in bulk diamond. We demonstrate cavity finesse at cryogenic temperatures within the range of $F = 4,000-12,000$ and find a sub-nanometer cavity stability. Modeling shows that coupling nitrogen-vacancy centers to these cavities could lead to an increase of remote entanglement success rates by three orders of magnitude.
\end{abstract}

\maketitle

Nitrogen-vacancy (NV) centers in diamond are promising building blocks for realizing quantum networks for computation, simulation and communication. The NV center electron spin and nearby nuclear spins form a robust multi-qubit quantum network node that is fully controlled by microwave and optical pulses\cite{Dutt,Tim}. Separate network nodes can be entangled through spin-photon entanglement and subsequent two-photon interference and detection \cite{Kok,Gao_Hannes,Bas_Bell}. The success rate of such entangling protocols is limited by the low probability (few percent) of the NV center emitting into the resonant zero phonon line (ZPL). Coupling of an NV center to an optical cavity can greatly increase the rate of generation and collection of ZPL photons through Purcell enhancement\cite{Purcell}. Purcell enhancement of the ZPL has been demonstrated in several cavity architectures such as diamond photonic crystal cavities\cite{Faraon_phc,Marko_phc,Hu_phc,Englund_phc,Becher_phc}, microring resonators\cite{Faraon_microring} and hybrid structures with evanescently coupled nanodiamonds\cite{Englund2010,Benson,Barclay2011,Toeno_phc}. In recent years the open Fabry-Perot microcavity \cite{Hunger2010} has emerged as a promising platform for diamond emitters\cite{Kaupp,Becher,JasonSmith,Lily,Kaupp_arxiv}. Such a microcavity provides \emph{in-situ} spatial and spectral tunability, while reaching strong field confinement due to its small mode volume $V$ and high quality factor $Q$. Moreover, this architecture allows for the use of diamond slabs\cite{Magyar_membranes} in which the NV center can be relatively far removed from surfaces and thus exhibit bulk-like optical properties, as required for quantum network applications.

Here we report on the realization of a high-finesse tunable microcavity enclosing a diamond membrane and its characterization under cryogenic conditions as relevant for quantum network applications. Our cavity employs a concave fiber tip fabricated using a CO$_{2}$ laser ablation technique\cite{Fiber_hunger} coated with a dielectric mirror stack, and a high reflectivity plane mirror onto which a thin diamond membrane is bonded (see Figure 1(a)). This cavity configuration is mounted inside a closed-cycle cryostation (Montana Instruments). To minimize scattering loss as required for a high finesse optical cavity, low surface roughness at the mirror-diamond and diamond-air interfaces is essential. We fabricate the diamond membrane (Figure 1(b)) by etching a polished \SI{30}{\micro m} thick diamond sheet (ElementSix) down to \SI{\approx 4}{\micro m} using Ar/Cl$_2$ inductively coupled plasma reactive ion etching. This etching process is known to preserve the surface smoothness of the diamond\cite{Enlund_Etch,Lee_Etch}. Using AFM, we measure a final diamond roughness value of 0.35 nm RMS. Finally the membrane is bonded to the plane mirror by Van der Waals forces\cite{Pawel}.

    \begin{figure}[t!]
		       \includegraphics[width=0.5\textwidth]{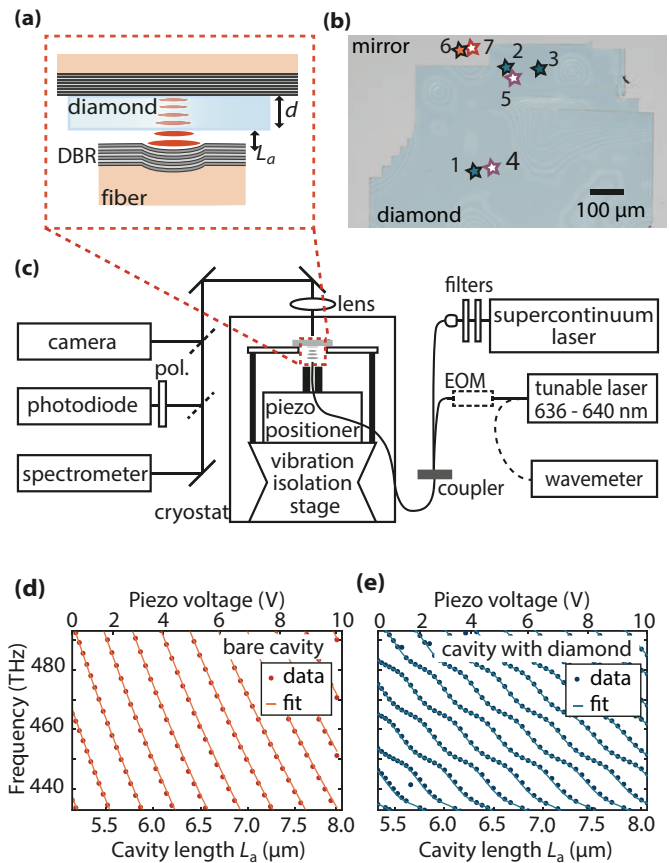}
        \caption{\label{fig1}(Color Online) Design of the setup and the cavity transmission spectra. (a) Schematic of the cavity showing the concave fiber-tip and the plane mirror onto which a diamond membrane is bonded. The fiber-tip concavity (radius of curvature of \SI{18.4}{\micro m}) is fabricated by CO$_2$ laser ablation and coated with a dielectric mirror (LASEROPTIK). The residual transmission of the fiber mirror is 50 ppm and losses are $\approx$ 70 ppm. The reflectivity of the plane mirror is $\approx 99.99\%$. From this follows an expected finesse of $F \approx 29,000$.
        (b)	Microscope image of the $4 \, \mu$m thick bonded diamond membrane. Positions at which cavity properties are measured at 300 K (11 K) are marked with filled (open) markers.
        (c)	Overview of the experimental setup. Effects of mechanical noise sources such as the cryostation pulse tube operation are mitigated with a high-frequency resonance cryo-positioning stage and a low-frequency resonance passive vibration isolation stage (Janssen Precision Engineering CPSHR1-s and CVIP1). See Supplementary Material for low temperature mechanical noise spectrum.
        (d),(e) The cavity fundamental modes dependency on the cavity length, for a bare cavity (position 6 in (b)) and a cavity containing diamond (position 1 in (b)). Higher order modes are removed to show only the fundamental mode which we fit with the resonant frequencies given by $\nu=cm/2 L_a$ for (d) and by Eq. 1 for (e).
}
    \end{figure}

We first study the cavity modes by recording transmission spectra as a function of cavity length using broadband excitation from a supercontinuum laser (see Figure 1(c)). From these spectra we extract the frequency of the fundamental modes of the cavity. The fiber mirror can be moved laterally to obtain an empty cavity (spectrum in Figure 1(d)), or a cavity including a diamond membrane (Figure 1(e)). The notably different length dependency for the two cases is a direct consequence of the presence of the high refractive index $(n_{d}=2.417)$ diamond membrane within the optical cavity. The partially reflecting interface between diamond and air creates a configuration in which the cavity field can be localized in air-like modes, with a length dependency similar to Figure 1(d), and in diamond-like modes, that are largely insensitive to changes in the cavity length. Due to the coupling between these modes, the behaviour of the fundamental modes in Figure 1(e) displays avoided crossings. The resulting resonant cavity frequencies $\nu$ are determined from a one-dimensional lossless cavity model\cite{Jayich2008, Lily}:
    \begin{equation}
    \begin{split}
    \nu\approx&\frac{c}{2\pi(L_a+n_{d}d)}\Bigg\{\pi m-(-1)^{m}\times\\
        & \arcsin \left(\frac{n_{d}-1}{n_{d}+1}\sin\left(\frac{m\pi(L_a-n_{d}d)}{L_a+n_{d}d}\right)\right)\Bigg\},
    \end{split}
    \end{equation}
and fit to the measured resonant frequencies to extract the diamond thickness $d$ and the length of the air layer in the cavity $L_a$ with an accuracy up to $\lambda/2$.

The intrinsic cavity properties are described by the finesse that we calculate using the cavity length (as determined by the transmission spectra) and the cavity linewidth in frequency. To measure the latter we couple light with a frequency of 471.3 THz from a narrow-linewidth ($< 1$ MHz) diode laser into the cavity and detect the transmitted signal using a photodiode as we scan the cavity length across the resonance. Phase-modulation was used to create laser sidebands at a fixed 6 GHz separation to directly determine the cavity linewidth in frequency (Figure 2(a,b)). We obtain the finesse of the cavity for different cavity lengths. These measurements are repeated at different positions on the diamond membrane and at different temperatures (300 K and 11 K). The results are summarized in Figure 2(c). For intermediate cavity lengths, high finesse values of approximately $10,000$ are supported by our cavity architecture. For cavity lengths larger than $55\times\frac{\lambda}{2}$, we observe a distinct drop in finesse which we attribute to clipping losses\cite{Hunger2010}. At short cavity lengths ($<45\times\frac{\lambda}{2}, \ L_{air}$ \SI{\approx 4}{\micro m}) the finesse values show significant fluctuations. We note that similar scatter of finesse values at short microcavity lengths has been previously observed\cite{Hunger2010,Lily}; potential causes are cavity misalignment and contact between the fiber and the plane mirror.

    \begin{figure*}[t!]
        \includegraphics[keepaspectratio=True]{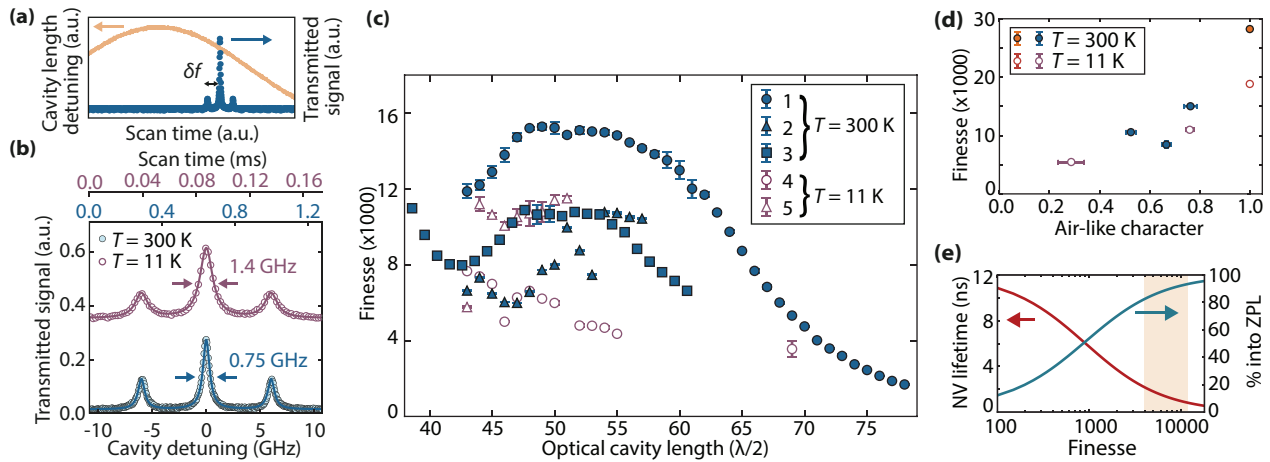}
        \caption{\label{fig2}(Color online) Measurements of intrinsic cavity properties. (a) Cavity linewidth measurements are performed by scanning the cavity length (orange) around the laser resonance and measuring the signal on the photodiode (blue). The laser frequency is modulated at $\delta f = 6$ GHz. (b) Two representative linewidth scans measured at $T = 300$ K and $T = 11$ K. A single polarization eigenmode is selected using a polarizer in the detection path. At cryogenic temperatures some scans show a deformation as a result of the system vibrations. To represent the intrinsic (vibration-independent) linewidth, we use only scans to which three Lorentzians could be reliably fitted.  (c) Finesse dependence on cavity length measured at five different positions on the diamond membrane at \emph{T} = 300 K (closed markers) and \emph{T} = 11 K (open markers). Per cavity length 40-100 scans as in (b) are averaged to obtain the linewidth in frequency. (d) Finesse dependence on the air-like character of the cavity mode, averaged over $L=47\times\frac{\lambda}{2}$ to $L=55\times\frac{\lambda}{2}$. The data points with an air-like character of 1 represent measurements of the bare cavity finesse. (e) Simulations of the excited state lifetime and emission probability into the cavity mode via the ZPL for an NV center embedded in this optical cavity with $L=45\times\frac{\lambda}{2}$. The shaded region shows the finesse range $4,000-15,000$ measured for cavities containing diamond.}
    \end{figure*}

We further investigate the variation of the average finesse as a function of the character (air-like versus diamond-like) of the cavity mode. Cavities formed at the steepest part of a mode (Figure 1(e)) are assigned an `air-like character' of 1, whereas the cavities at the flat part have air-like character of 0. Intermediate values are obtained from a linear interpolation by frequency. The bare cavity, that we approximate to have an air-like character of 1, has a finesse of $\emph{F}\approx 28,000$ (Figure 2(d)), which is in agreement with the value expected from the mirrors' parameters. Inserting the diamond membrane into the cavity reduces the finesse. We attribute this reduction to several effects. First, adding a diamond interface into the optical cavity introduces an additional loss mechanism due to scattering from the diamond surface. Given the measured surface roughness of the diamond membrane we expect a reduction in finesse due to scattering to $\emph{F}\approx 21,000$\cite{Scattering}. Second, the refractive index of the plane mirror coating is optimized for bare cavity applications. Inserting a diamond membrane (which has a higher refractive index than air) will lower its effective reflectivity, reducing the finesse threefold\cite{Lily}. The influence of these mechanisms is strongly dependent on the character of the mode in the cavity. The modes with a diamond-like character have an antinode at the air-diamond interface, and therefore are most susceptible to scattering at the diamond surface. The trend in the data in Figure 2(d) is consistent with the above consideration, where modes with a more air-like character show a higher finesse.

We estimate the effect that the cavities realized here would have on an embedded NV center's excited state lifetime as well as the probability that emission occurs via the ZPL into the cavity mode (Figure 2(e)). We use the Purcell factor $F$ for an ideally placed and oriented NV:
\begin{equation}
F=\frac{3}{4\pi^2}\left(\frac{c}{n_d\nu}\right)^3\frac{Q}{V},
\end{equation}
and use bulk-like free-space values for the branching ratio into the ZPL (3\%) and excited state lifetime (12 ns)\cite{Faraon_microring}. A more complex model that explicitly takes dephasing, phonon side-band emission and other cavity modes into account \cite{Auffeves,Becher} yields quantitatively similar results (not shown).
We find that the emission properties of the NV center would be greatly improved, with a probability of emission into the cavity mode via the ZPL above $80\%$ for the current finesse values, compared to the $\approx 3\%$ probability into all modes for the uncoupled case. Thus, both the relative contribution of ZPL photons to the emission as well as the collection efficiency may be significantly enhanced using these cavities.

    \begin{figure}[t!]
        \includegraphics[width=0.5\textwidth]{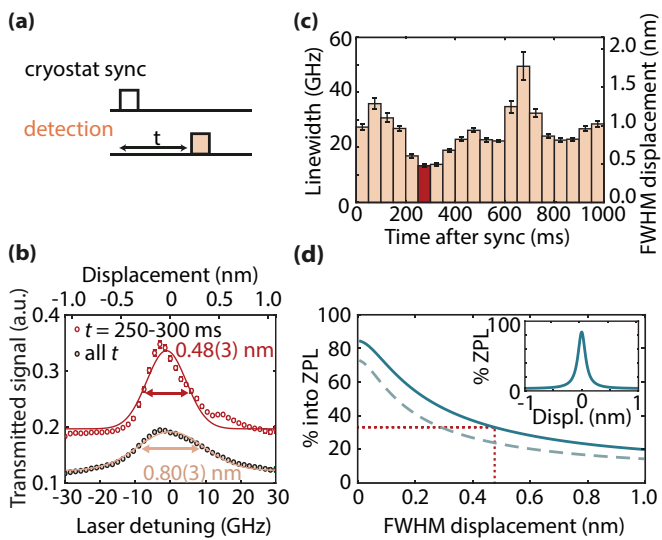}
        \caption{\label{fig3}(Color online) Vibration-sensitive measurements of the cavity linewidth. (a) Timing of the cavity linewidth detection with respect to the cryostation synchronization signal. (b) Measurement of the cavity transmitted signal, performed by sweeping the laser frequency over the cavity resonance during 41 cycles of the cryostat pulse tube. The center of 50 sweeps is overlapped and averaged, and fitted with a Gaussian curve, for data collected throughout the cryostation tube cycle (orange curve), and for data collected in the time bin 250-300 ms after the sync signal (red curve). (c) Cavity linewidth dependency on the measurement time with respect to the sync signal. (d) Simulation of the NV center emission via the ZPL for a cavity with length $45\times\frac{\lambda}{2}$ subject to vibrations. The results include a perfectly oriented emitter in the cavity anti-node (solid line) and for an emitter with 30$^{\circ}$ dipole mismatch and $\frac{\lambda}{10}$ deviation of the emitter position from the cavity anti-node (dashed line). The inset shows the dependency of the NV center's emission into the ZPL on the cavity displacement from its resonance position.}
    \end{figure}

The linewidth measurements in Figure 2 probe the intrinsic cavity properties at time scales comparable to the scan time (0.1 ms at $T=11$ K). Cooling the system to cryogenic temperatures introduces significant low-frequency (up to about 10 kHz) mechanical noise from the cryostation pulse tube, which results in cavity linewidth broadening when averaging over time scales longer than (10 kHz)$^{-1}$. We probe the effect of the low-frequency noise on the system by measuring the cavity transmission signal as a function of the laser frequency at a fixed cavity length  ($50\times\frac{\lambda}{2}$). The laser frequency is swept slowly compared to  the pulse tube cycle time, ensuring that the full effect of pulse-tube-induced vibrations is visible in the data. The resulting signal is shown in the orange curve in Figure 3(b). The broadened cavity linewidth is fitted with a Gaussian function, for which a full width half maximum (FWHM) of \SI{22.2+-.7}{GHz} is found. This value is a direct measure of the cavity displacement from its resonance position of \SI{0.80+-0.03}{nm}.

Synchronization of our measurement to the 1-Hz cycle of the cryostation pulse tube gives further insight into the effect of the mechanical noise. In Figure 3(c) we present the dependence of the effective cavity linewidth on the measurement delay with respect to the cryostation sync signal (Figure 3(a)). We find that the vibrations of the system are strongly dependent on the timing within the cryostation cycle, with the cavity linewidth broadening varying from 14 GHz to 50 GHz. The open red datapoints in Figure 3(b) show the photodiode signal for the lowest vibration time-bin, 250-300 ms after the sync signal, for which the Gaussian fit gives a cavity length displacement of \SI{0.48+-0.03}{nm}. Cavity displacement can be further reduced by employing active cavity stabilization methods such as the Pound-Drever-Hall technique \cite{Black2001}.

 Figure 3(d) shows the effect of the low-frequency vibrations on the expected fraction of the NV center's emission into the ZPL as calculated in Figure 2(d). We use a Gaussian distribution of the displacements as found in the vibration-sensitive measurement of Figure 3(b) and a target cavity finesse of 5,000. For the measured vibration levels, we expect the resulting emission via the ZPL into the cavity mode to be 33\% which still greatly surpasses the native NV center's emission. In the analysis, we assume the case of an ideally placed emitter within the cavity field (Figure 3(d) (solid line)). We additionally explore the effect of a non-ideal dipole orientation and emitter location, resulting in an emission probability of 26\% (Figure 3(d) (dashed line)). In practice, close-to-ideal conditions could be achieved by utilizing a $\langle111\rangle$-oriented diamond crystal and achieving a high NV-center concentration through nitrogen implantation \cite{Chu} or nitrogen delta-doping growth \cite{Ohno}. Stable implanted NV centers with the desired linewidths have already been reported\cite{Chu}.

In conclusion, our tunable, high-finesse Fabry-Perot microcavity with an embedded diamond membrane reaches high finesse values of $\emph{F} \approx 12,000$ at cryogenic temperatures. The demonstrated 0.48~nm length stability under these conditions would enable an approximately 13 times increase in the NV ZPL photon emission. Additionally, these resonant photons are all fed into the well-defined spatial cavity mode that is well suited for collection, leading to an estimated $3$ times enhanced collection efficiency. For demonstrated NV center remote entangling schemes that rely on two-photon interference\cite{Kok,Bas_Bell} the resulting boost in the generation and collection of resonant photons in the presented architecture would thus offer an $(3\times13)^2 \approx 10^3$ increase in success probability.\\

\emph{Supplementary Material}: See Supplementary Material for the cavity noise spectral properties.\\

The authors wish to thank P. Latawiec, L. Childress and E. Janitz for helpful discussions. M.L. wishes to acknowledge the support of QuTech during his sabbatical stay. M.S.Z.L. acknowledges the Dutch Liberation Scholarship Programme. This work was supported by the Dutch Organization for Fundamental Research on Matter (FOM), Dutch Technology Foundation (STW), the Netherlands Organization for Scientific Research (NWO) through a VICI grant, the EU S3NANO program and the European Research Council through a Starting Grant.

\bibliography{Bogdanovic_manuscript}
\end{document}